\documentclass{article}

\usepackage[english]{babel} 
\usepackage[utf8]{inputenc} 
\usepackage[a4paper,left=3cm,bottom=3cm]{geometry} 
\usepackage{indentfirst} 
\usepackage{enumerate, enumitem} 
\usepackage{dsfont, mathtools, amssymb, amsmath} 
\usepackage{bm, booktabs} 
\usepackage{graphicx} 
\usepackage{multicol} 
\usepackage[export]{adjustbox} 
\usepackage{pdflscape} 
\usepackage{fancyhdr} 
\usepackage{hyperref}
\hypersetup{colorlinks,allcolors=blue}
\usepackage{flafter} 
\usepackage[framemethod=tikz]{mdframed} 
\usepackage{color} 
\usepackage{wrapfig}
\usepackage{float} 
\usepackage{lipsum}
\usepackage{xcolor, xurl}

\graphicspath{ {Images/} }  

\title{A Predictive Interference Management Algorithm for URLLC in Beyond 5G Networks}
\author{Nurul H. Mahmood, Onel A. L\'{o}pez, Hirley Alves and~Matti Latva-aho
\thanks{The authors are with 6G Flagship, Centre for Wireless Communications, University of Oulu, Finland. Corresponding e-mail: \textit{nurulhuda.mahmood@oulu.fi}. This work is supported by the Academy of Finland 6Genesis Flagship program (grant no. 318927). \textit{Submitted to IEEE for possible publication.}}
}

\date{\today}

\begin{document}

\pagenumbering{gobble} 
\maketitle

\pagenumbering{arabic} 
\setcounter{page}{1}


\begin{abstract}
Interference mitigation is a major design challenge in wireless systems,especially in the context of ultra-reliable low-latency communication (URLLC) services. Conventional average-based interference management schemes are not suitable for URLLC as they do not accurately capture the tail information of the interference distribution. This letter proposes a novel interference prediction algorithm that considers the entire interference distribution instead of only the mean. The key idea is to model the interference variation as a discrete state space discrete-time Markov chain. The state transition probability matrix is then used to estimate the state evolution in time, and allocate radio resources accordingly. The proposed scheme is found to meet the target reliability requirements in a low-latency single-shot transmission system considering realistic system assumptions, while requiring only $\sim 25\%$ more resources than the optimum case with perfect interference knowledge. \newline
\textit{\textbf{Keywords: }Beyond 5G/6G networks, intelligent resource management, interference prediction, URLLC.}
\end{abstract}

\section{Introduction}
\label{sec:introduction}
Ultra-reliable low-latency communication (URLLC) is a novel service class introduced in the latest fifth generation (5G) New Radio (NR) wireless network standard. URLLC targets stringent reliability performances, e.g., in the order of $99.999\%$, and latency budgets as low as one millisecond~\cite{SWD+18}. These stringent quality of service (QoS) figures are required to support a variety of mission-critical applications in different vertical industries, such as remote control of robots, autonomous coordination among vehicles and in industrial automation~\cite{berardinelli2018_wirt}.

There are three main lower layer challenges in ensuring URLLC, namely \textit{(i)} uncertainty in the traffic arrival, \textit{(ii)} channel impairments such as fading, and \textit{(iii)} random interference. Addressing these challenges efficiently mandate a departure from the average utility based approach of conventional radio resource management (RRM) practices to a framework that considers the tail behavior of the reliability, latency and throughput performance~\cite{Bennis.2018}.

URLLC enablers in 5G NR are primarily centered around redesigning the system numerology to meet low latency requirements~\cite{SWD+18}, and ensuring high reliability through over-provisioning of resources~\cite{MLP+19_MCA}. While such an approach is known to enable URLLC under sparse and controlled environments, it is neither scalable nor resource efficient, and therefore does not adequately address the fundamental and novel challenges imposed by URLLC. Alongside, emerging URLLC use cases in future beyond 5G/sixth generation (6G) networks will require QoS guarantees that are much more diverse and stringent than those considered in 5G NR~\cite{MTCwhitePaper2020}. Novel solutions incorporating intelligent and predictive RRM algorithms are therefore needed to enable URLLC in a scalable and resource efficient manner in future wireless networks. 

Link adaptation is a well known RRM approach in conventional interference management. It involves estimating future interference values from past samples, which is then mitigated by adapting the link parameters accordingly~\cite{Pocovi.2017}. Conventional link adaptation schemes operate by condensing the entire interference distribution into a single mean value, and are therefore not suitable for URLLC applications where the tail behavior of the interference distribution needs to be considered~\cite{Bennis.2018}. Instead, interference prediction strategies that capture information about the entire interference distribution are better suited. Generally, such interference prediction involves building a model to reflect the interference variation across time, which is then used to predict future interference values. 

RRM algorithm design for URLLC applications based on interference prediction is still at its infancy (see for example~\cite{LCH+20}). We aim at contributing to this emerging field by proposing a novel interference prediction based RRM algorithm in this work. We model the interference variation as a discrete state space discrete-time Markov chain (DTMC). The state transition probability matrix, obtained by observing the history of state transitions, is used to predict future interference states. Finally, the predicted interference is mitigated through efficient allocation of resources. The proposed approach is validated using extensive Monte-Carlo simulations. We observe that very low outage probabilities can be achieved with the proposed scheme, which are otherwise not possible by adopting a conventional average-based approach.


\section{System Model}
\label{sec:systemModel}
We consider the downlink\footnote{The proposed scheme is link direction agnostic and works equally for the uplink as well.} of a wireless network. The focus is on a desired URLLC link operating in the presence of $N$ interferers distributed in $\mathbb{R}^2$ space. The desired channel is assumed to have a mean signal to noise ratio (SNR) of $\bar{\gamma}_{D}$, whereas the mean interference to noise ratio (INR) corresponding to each interfering link - i.e., the mean SNR of the interference signal - is considered to be uniformly distributed in the range $[\bar{\gamma}_{I,min}, \bar{\gamma}_{I,max}].$ We assume $\bar{\gamma}_{D} \geq \bar{\gamma}_{I,max},$ i.e., users are served by the transmitter with the strongest mean SNR. All nodes are assumed to operate independently, i.e., there is no cooperation among them. A single-antenna Rayleigh block-fading channel model is adopted. 

The desired transmitter transmits a short packet of $D$ bits with a target outage/block error rate (BLER) $\epsilon.$ The transmitter estimates the signal to interference plus noise ratio (SINR), and then allocates the required resources accordingly. The goal is to estimate the SINR as accurately as possible to ensure efficient resource allocation.

URLLC aims to realize low latency and high reliability simultaneously. This is usually done by analyzing the achieved reliability considering a given latency budget~\cite{MAB+19_grantFree}. In this work, we consider a very tight latency budget that cannot accommodate any retransmission, i.e., transmissions are assumed to be single shot~\cite{MPJ+16_oneStage}. Such an assumption allows us to analyze the lower bound of URLLC performance since retransmissions improve reliability, albeit at the cost of additional latency~\cite{Popovski.2018}. 

URLLC transmissions usually occur over mini-slots of duration $\sim 0.1$ ms, thereby providing shorter and more agile transmission units compared to the $1$ ms transmission time interval (TTI) of LTE~\cite{SWD+18}. Given that the channel coherence time in a typical wireless environment is much larger than the mini-slot duration, the duration of a fading block for URLLC transmissions spans over multiple TTIs. The desired transmitter is therefore assumed to have sufficient time to acquire accurate knowledge of the desired channel state information (CSI). In addition, the receiver is also assumed to feed back the experienced `aggregate' interference allowing the transmitter to acquire knowledge of the past interference samples.


\section{The Proposed Interference Prediction Scheme}
\begin{figure*}
    \centering
    \includegraphics[width=0.99\linewidth]{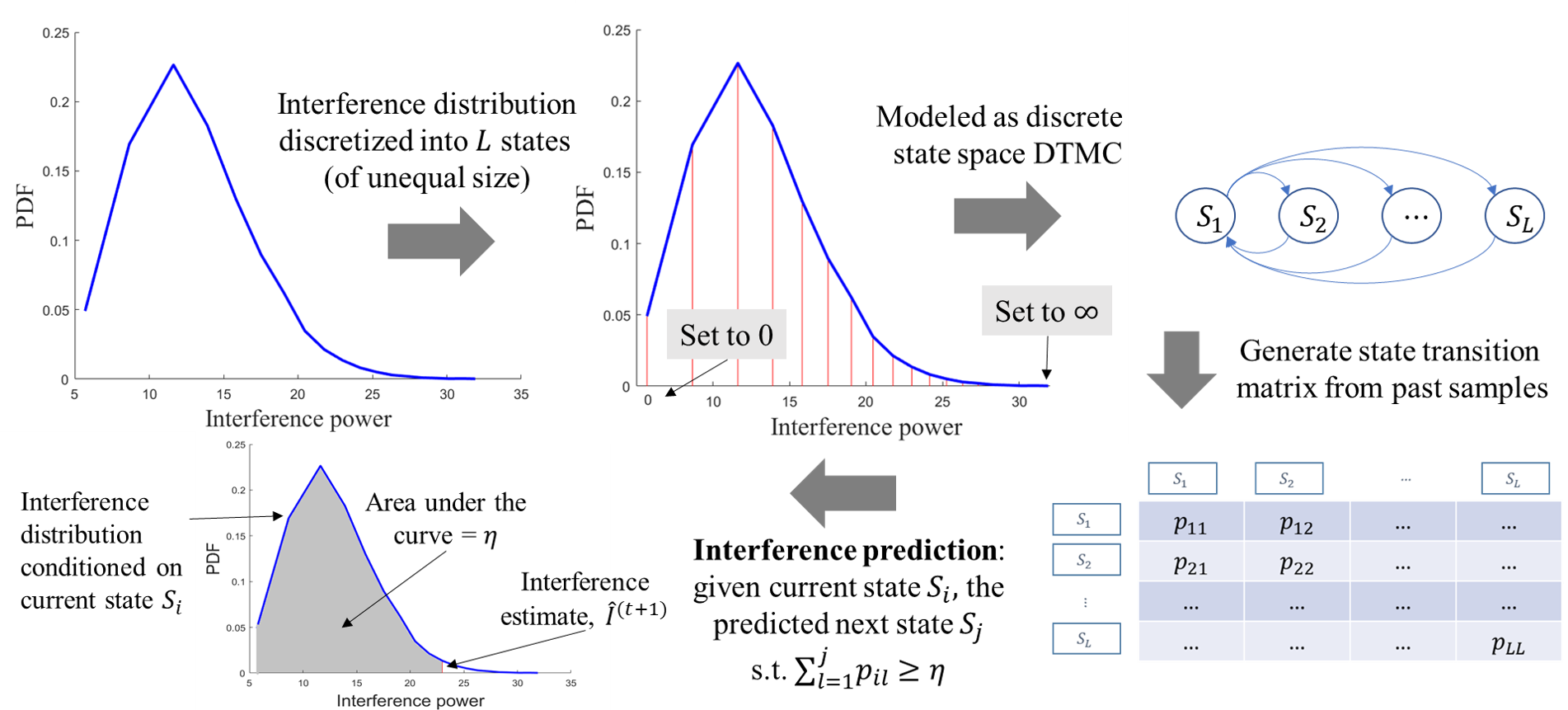}
    \caption{Sketch of the proposed algorithm.}
    \label{fig:outline}
\end{figure*}

A model-based approach to interference prediction is adopted in this work, where the interference distribution is modeled as a discrete state space DTMC. The sketch of the proposed interference prediction algorithm is pictorially outlined in Fig.~\ref{fig:outline} and detailed in the rest of this section. 

\subsubsection{Discretization of the Interference Space}
As the first step, the initially observed (continuous) interference space $\mathcal{I}$ is discretized into the state space $\mathcal{L}\triangleq \{\mathcal{S}_1, \ldots, \mathcal{S}_L\}$ such that interference values in the range $[\mathbb{I}_{l-1}, \mathbb{I}_l)$ are assigned to state $\mathcal{S}_l$. A straightforward option would be to discretize $\mathcal{I}$ into $L$ equally spaced states within the range of $\mathcal{I}$. However, this treats stronger and weaker interference values equally and is contradictory to the risk-sensitive approach recommended for URLLC~\cite{Bennis.2018}. We therefore discretize $\mathcal{I}^2$ into $L$ equally spaced states and define the state boundaries of $\mathcal{L}$ to be the squareroots of those values. Thus, larger interference values are represented by a higher number of states, as depicted in Fig.~\ref{fig:outline}. Finally, we set $\mathbb{I}_0 = 0$ and $\mathbb{I}_L = \infty$ to capture any future interference value that may lie outside the range of $\mathcal{I}$. It is worth mentioning that, the choice of $L$ results in a trade-off between algorithm complexity and performance accuracy.

\subsubsection{Obtaining the Transition Matrix}
Next, we obtain the transition matrix $\mathcal{P}$ describing the transition probabilities of the states in $\mathcal{L}$. We first define $\tilde{\mathcal{I}} \triangleq \{\mathcal{S}^{0}, \mathcal{S}^{1}, \ldots \}$ as the set of observed interference states at time index $t = 0, 1, \ldots$. The transition probability $p_{ij}$ denotes the probability of a transition from the states $\mathcal{S}_i$ to $\mathcal{S}_j$, and is defined as 
\begin{equation}
    p_{ij} = \frac{\sum_t \left[\mathds{1}_{\mathcal{S}^{t+1}}\left(\mathcal{S}_j\right)
    | \mathds{1}_{\mathcal{S}^{t}}\left(\mathcal{S}_i\right) \right]}
    {\sum_t \left[\mathds{1}_{\mathcal{S}^{t}}\left(\mathcal{S}_i\right) \right]},
    \label{eq:transitionProbability}
\end{equation}
where $\mathds{1}_{\mathcal{A}}(x)$ is the indicator function which equals $1$ if $x \in \mathcal{A}$ and $0$ otherwise. All the entries of the transition matrix $\mathcal{P}$ are thus obtained by evaluating~\eqref{eq:transitionProbability} $\forall \, i, j \in \{1, 2, \ldots, L\}.$

\subsubsection{Interference Prediction}
In order to utilize the tail statistics of the interference distribution, we introduce the confidence level parameter $\eta < 1$, which can be viewed as a \textit{risk-sensitivity index}~\cite{Bennis.2018}. Suppose, we would like to utilize the right end of the distribution tail up to a point $\zeta$. $\eta$ is then the area under the left side of the PDF up to point $\zeta$. Hence, the closer $\eta$ approaches one, the closer to the right end of the distribution tail is utilized for the prediction, as illustrated in Fig.~\ref{fig:outline}. More specifically, $\eta$ characterizes the likelihood that the predicted interference level is higher than the actual interference. Thus, a larger $\eta$ corresponds to a more conservative interference estimation. 

Suppose, $\mathcal{S}^{t} = \mathcal{S}_i$, i.e., the interference state at time $t$ is $\mathcal{S}_i$. Our proposed scheme predicts the interference at time $t+1$ to be such that the \textit{predicted interference} $\left(\hat{I}^{t+1}\right)$ is greater than the \textit{actual interference} $\left({I}^{t+1}\right)$ with probability $\geq \eta.$ Mathematically, this is stated as $Pr[{I}^{t+1} \leq \hat{I}^{t+1}] \geq \eta$. To ensure this, we first predict the next state $\hat{\mathcal{S}}^{t+1}$ to be $\mathcal{S}_j$, where $j$ is the smallest integer such that $\sum_{l=1}^{j} p_{il} \geq \eta$. The predicted interference level at time index $t+1$ is then
\begin{align}
    \hat{I}^{t+1} = 
    \begin{cases}
        \mathbb{I}_j & \text{if } j \neq L \\
        2\mathbb{I}_{L-1} - \mathbb{I}_{L-2} & \text{if } j = L
    \end{cases},
    \label{eq:predictedInterf}
\end{align}
where $\mathbb{I}_j$ is the right-endpoint of state $\mathcal{S}_j$. Please note that, we have earlier set $\mathbb{I}_L = \infty$ and hence if $\hat{\mathcal{S}}^{t+1} = \mathcal{S}_L$, we risk obtaining $\hat{I}^{t+1} = \infty.$ To avoid this, the second step in~\eqref{eq:predictedInterf} sets a \textit{dummy} right-endpoint of $\mathcal{S}_L.$ Note that the \textit{dummy} endpoint can be set to any value greater than $\mathbb{I}_{L-1}$ for the proposed scheme to work.  

\subsubsection{Transition Matrix Update}
The last step of the proposed algorithm involves updating the transition matrix $\mathcal{P}$. Suppose, a transition is made from state $\mathcal{S}_i$ to state $\mathcal{S}_j. \mathcal{P}$ is then updated as follows,
\begin{align*}
	&\text{\textit{step 1, update:}}\ p_{ij} \rightarrow p_{ij} + \omega_{ij} \\
	&\text{\textit{step 2, normalize:} normalize $i^{th}$ row s.t. }\sum_{j=1}^{L} p_{ij} = 1.
\end{align*}
Here, $0 \leq \omega_{ij} \ll 1$ is the learning rate. The provision for having different learning rates for different states allows giving a higher weight to certain transitions. In this work, we have set $\omega_{ij} = \omega_{i}\ \forall j$ and let its value be inversely proportional to the number of elements in state $\mathcal{S}_i.$ 

\subsubsection{Complexity Analysis}
The proposed interference prediction algorithm is of low-complexity with minimal additional signaling overhead. The only additional step during runtime involves looking up the transition matrix (a table of size $L$) to predict the next interference state. In addition, the algorithm requires the receiver to feed back the experienced interference level along with the CSI, and the transmitter to update $\mathcal{P}$ upon receiving this information. All of these steps are rather simple to execute and incurs negligible processing delay. 


\section{Numerical Evaluation}
\label{sec:results}

The performance of the proposed scheme is evaluated and compared against two baseline schemes in this section. 

\subsection{Baseline Estimation Techniques}
\label{sub:baseline}

\subsubsection{Moving Average Based Estimation}
The conventional weighted average based interference estimator, which is adopted as an estimator in link adaptation for conventional enhanced mobile broadband (eMBB) services~\cite{Pocovi.2017}, is considered as the first baseline scheme. In this scheme, the interference measurement at time $t$ is filtered with a low-pass first-order Infinite Impulse Response (IIR) filter, resulting in the following interference estimate, 
\begin{equation}
	\hat{I}_{t+1} = \alpha I_{t-1} + (1 - \alpha)\hat{I}_{t},
\end{equation}
where $0 < \alpha < 1$ is the forgetting factor (FF) of the filter. 

\subsubsection{Genie Aided Estimation}
We also consider a genie-aided estimator where the exact interference condition is known \textit{a priori} at the transmitter, leading to optimum resource allocation. Even though such a scheme is not feasible in practice, it is considered as an indicator of the optimum performance bound. 

\subsection{Resource Allocation}
\label{sub:resourceAllocation}
Let $\gamma = {\sigma}/{\hat{I}}$ be the predicted SINR, where $\sigma$ is the SNR of the desired transmission (which we assume to be known using CSI estimates) and $\hat{I}$ is the predicted interference level as described in the preceding section. Using results from finite blocklength theory, the number of information bits $D$ that can be transmitted with decoding error probability $\epsilon$ in $R$ channel uses in an additive white Gaussian noise (AWGN) channel with SINR $\gamma$ is given as~\cite{polyanskiy_trIT2010}
\begin{align}
	\label{eq:polyanskiy}
	D = R C(\gamma) - Q^{-1}(\epsilon)\sqrt{R V(\gamma)} + \mathcal{O}(\log_2 R),
\end{align}
where $C(\gamma) = \log_2 (1 + \gamma)$ is the Shannon capacity of AWGN channels under infinite blocklength regime, $V(\gamma) = \frac{1}{\ln(2)^2} \left( 1 - \frac{1}{\left( 1 + \gamma \right)^2}\right)$ is the channel dispersion (measured in squared information units per channel use) and $Q^{-1}(\cdot)$ is the inverse of the Q-function. Using the above, the channel usage $R$ can be approximated as~\cite{AV_jsac2018}
\begin{equation}
	\label{eq:channelUsageAnand}
	R \approx \frac{D}{C(\gamma)} + \frac{Q^{-1}(\epsilon)^2 V(\gamma)}{2 C(\gamma)^2} \left[ 1 + \sqrt{1 + \frac{4 D C(\gamma)}{Q^{-1}(\epsilon)^2 V(\gamma)}}\right].	    
\end{equation}

Eq.~\eqref{eq:polyanskiy} is extended in~\cite{LKD20_FBCsingle} to derive the achievable maximum coding rate spanning over multiple coherence intervals considering the more practical (and assumed) case of correlated Rayleigh block-fading channels. However, this only changes the absolute performance but not the relative performance measures of the different schemes, and hence does not alter the key findings of this paper. 

\subsection{Performance Evaluation}
\label{sub:perfEvaluation}

The performance of the proposed interference prediction method is numerically evaluated against the two baseline schemes presented in Section~\ref{sub:baseline}. Unless stated otherwise, the simulation parameters presented in Table~\ref{tab:simParameters} are adopted.

\begin{table}[t!]
\begin{center}
\caption{General simulation parameters}
\label{tab:simParameters}
\begin{tabular}{l c}
\toprule
Parameter & Value\\
\midrule
Mean SNR, $\bar{\gamma}_{D}$ 	& $20$ dB \\
Mean INR range, $[\bar{\gamma}_{I, min}, \bar{\gamma}_{I, max}]$ & $[-10, 5]$ dB \\
Number of interferers, $N$ 	& $5$ \\
Channel model 	& Rayleigh fading SISO \\
Packet length, $D$ 	& $50$ bits \\
Number of states, $L$ & $20$ \\
Risk sensitivity index, $\eta$ 	& $0.95$ \\
FF of IIR filter, $\alpha$ 	& $0.01$ \\
\bottomrule
\end{tabular}
\end{center}
\vspace{-6mm}
\end{table} 

\subsubsection{Outage Probability vs. Resource Usage}
\begin{figure}[h]
    \centering
    \includegraphics[width=0.65\columnwidth]{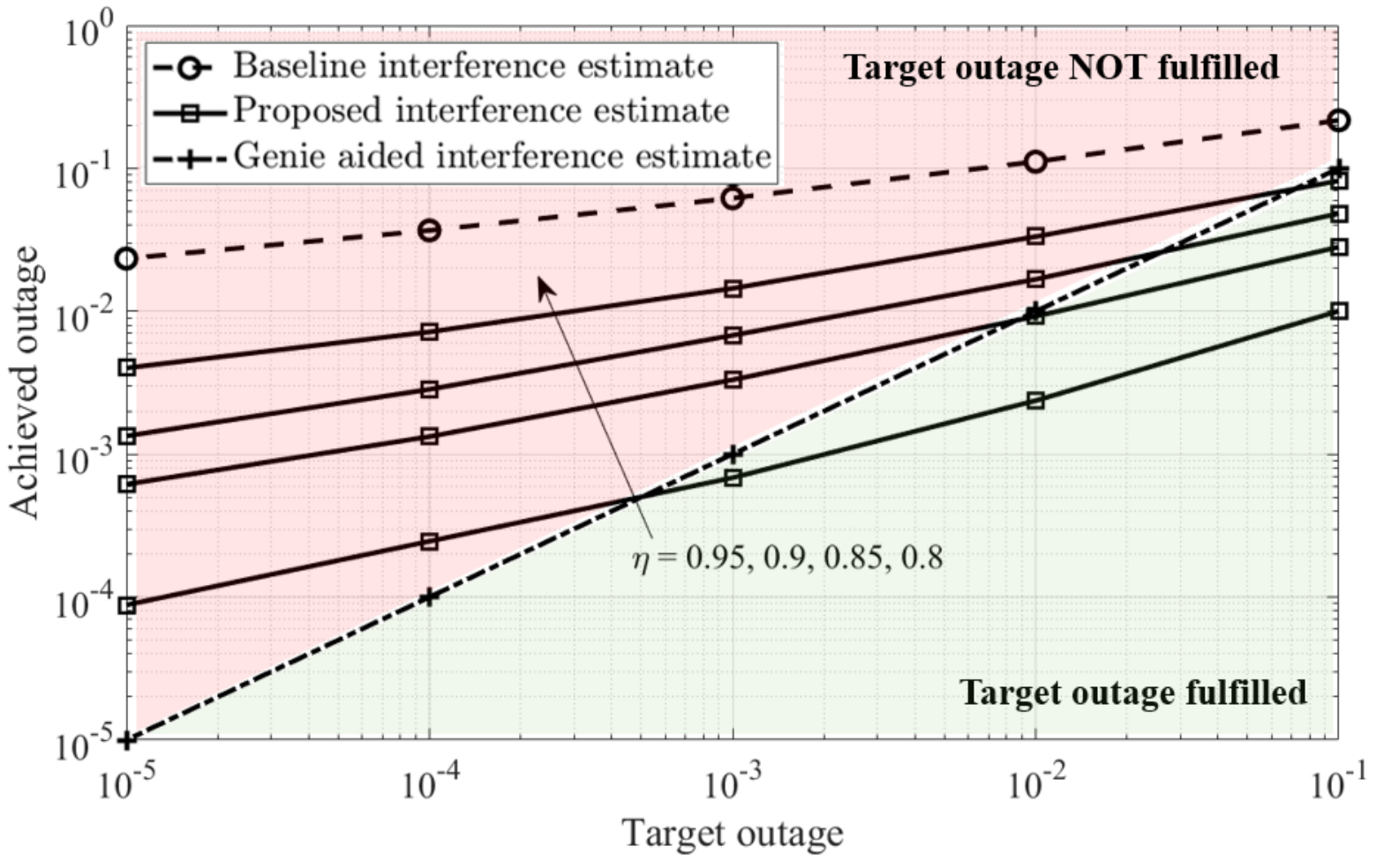}
    \caption{Achieved outage as a function of the target outage probability for different values of the risk-sensitivity index $\eta$.}
    \label{fig:outage_vs_eta}
\end{figure}

We first evaluate the outage probability as a function of the target outage $\epsilon$ as plotted in Fig.~\ref{fig:outage_vs_eta}. Four different values of the risk-sensitivity index are considered, namely $\eta = \{0.8, 0.85, 0.9, 0.95\}.$ Since the genie-aided estimator is assumed to know the achieved SINR beforehand, it can allocate the exact amount of resources required to meet the target outage. On the other hand, the conventional IIR filter based estimator performs very poorly and is only able to meet a BLER target of about $10\%.$ It is worth mentioning that the conventional IIR filter based approach is known to be quite resource efficient for eMBB services, where the BLER target is usually around $10\%$. 

The performance of the proposed scheme depends on $\eta.$ Outage targets as low as $5\times 10^{-4}$ can be fulfilled with a conservative value of $\eta = 0.95$.  Obviously, this comes at the cost of higher resource usage (RU), as discussed next. It is worth highlighting that this outage is achieved with a single shot transmission, which means that the outage performance can be significantly improved with retransmission or other diversity techniques, albeit at the cost of higher latency.

\begin{figure}[h]
    \centering
    \includegraphics[width=0.65\columnwidth]{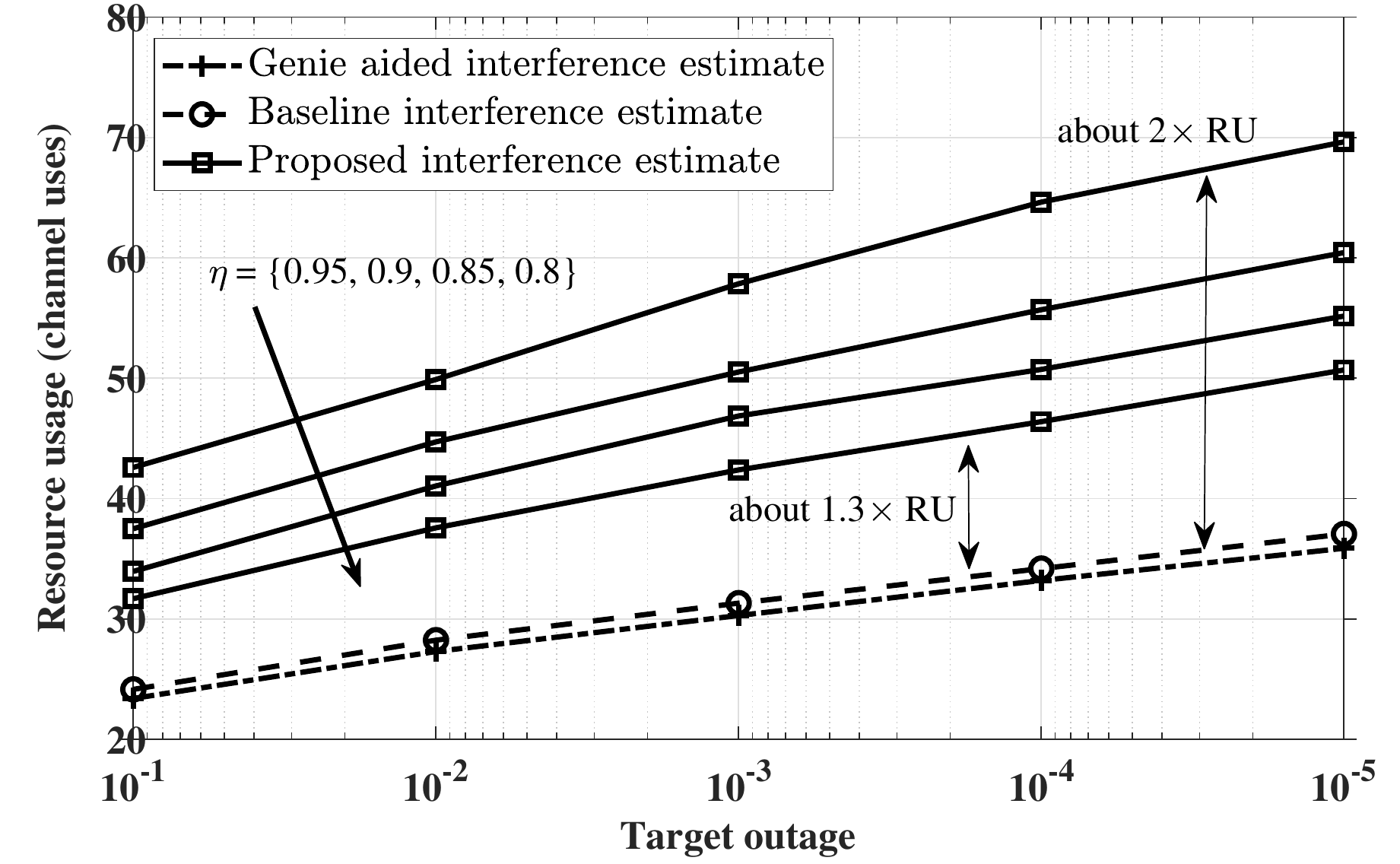}
    \caption{Resource usage in channel uses vs. target outage probability for different values of the risk-sensitivity index $\eta$.}
    \label{fig:outage_vs_RU}
\end{figure}

The average RU corresponding to the above setup is shown in Fig.~\ref{fig:outage_vs_RU}. As expected, the lowest RU is achieved when the SINR is exactly known \textit{a priori} as it leads to the optimum resource allocation. The IIR filter based baseline scheme is very close to the optimum performance in terms of RU. As indicated earlier, this is the conventional scheme in the case of eMBB services where optimum RU is prioritized over high reliability and low latency. 

The RU of the proposed scheme depends on the risk-sensitivity index $\eta$, which can be used to balance the trade-off between reliability and RU. A higher $\eta$ reflects a more conservative prediction. This leads to a lower SINR estimation, and subsequently higher resource allocation while also resulting in a better outage performance. 

\subsubsection{Performance Under Different Interference Conditions}
The performance of our proposed scheme under different interference conditions is evaluated in this section. We consider three different scenarios, as follows
\begin{description}[style=unboxed]
    \item[Strong SNR, strong interference:] Mean SNR and the INR range set to $\bar{\gamma}_{D} = 20$ dB, and $[\bar{\gamma}_{I, min}, \bar{\gamma}_{I, max}] = [0, 20]$ dB, respectively;
    \item[Strong SNR, weak interference:] $\bar{\gamma}_{D} = 20$ dB and $[\bar{\gamma}_{I, min}, \bar{\gamma}_{I, max}] = [-5, 5]$ dB; and 
    \item[Weak SNR, weak interference:] $\bar{\gamma}_{D} = 5$ dB and $[\bar{\gamma}_{I, min}, \bar{\gamma}_{I, max}] = [-5, 5]$ dB.
\end{description}

\begin{figure}[h]
    \centering
    \includegraphics[width=0.65\columnwidth]{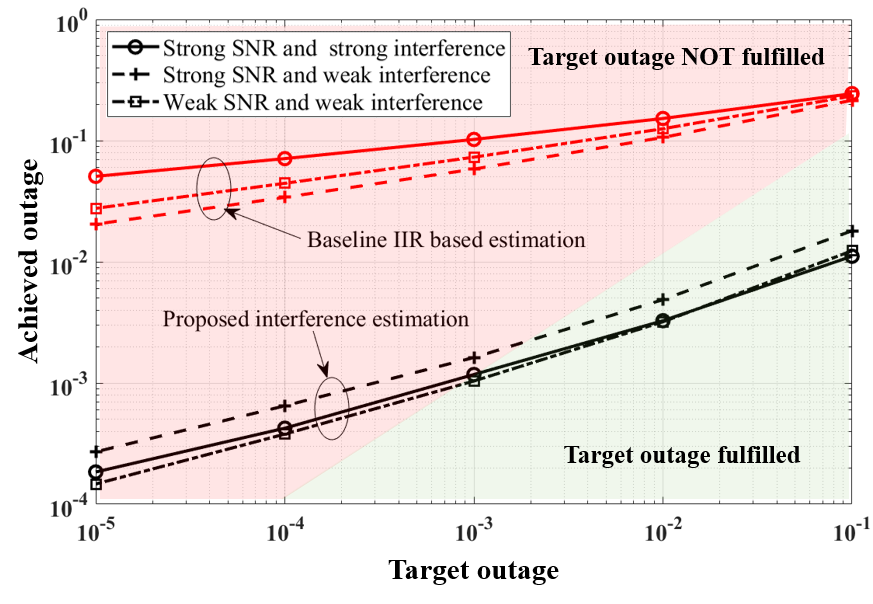}
    \caption{Outage probability vs. target outage under different interference scenarios, $\eta = 0.95$.}
    \label{fig:outage_diffScen}
\end{figure}

Fig.~\ref{fig:outage_diffScen} presents the outage probability as a function of the target outage under the considered interference scenarios. As a general trend, neither the baseline nor the proposed scheme is strongly sensitive to the interference scenario. The baseline scheme behaves as expected with the \textit{strong SNR, weak interference} scenario demonstrating the best performance, followed respectively by \textit{weak SNR, weak interference} and \textit{strong SNR and strong interference} scenarios. On the other hand, given its nature, the proposed interference prediction scheme can better track the interference when it is strong. Hence, a better performance is observed for the scenarios where the interference strength is comparable to the SNR.

\subsubsection{Performance Evaluation with Correlated Traffic}
In order to meet the low latency requirements, URLLC transmissions in 5G NR take place over mini-slots with rather short TTIs~\cite{SWD+18}. Hence, the channel coherence time in practice spans over multiple transmission slots, resulting in the interference from a given transmitter being correlated over time. 

\begin{figure}[h!]
    \centering
    \includegraphics[width=0.65\columnwidth]{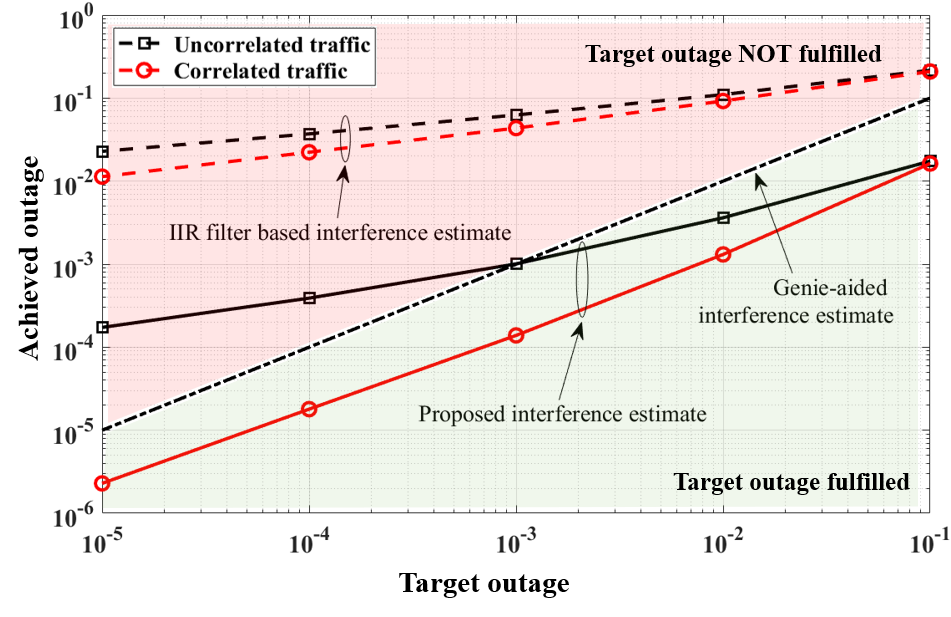}
    \caption{Achieved outage as a function of the target outage probability with correlated traffic, $\eta = 0.95$.}
    \label{fig:correlated}
\end{figure}

\begin{figure}[h!]
    \centering
    \includegraphics[width=0.65\columnwidth]{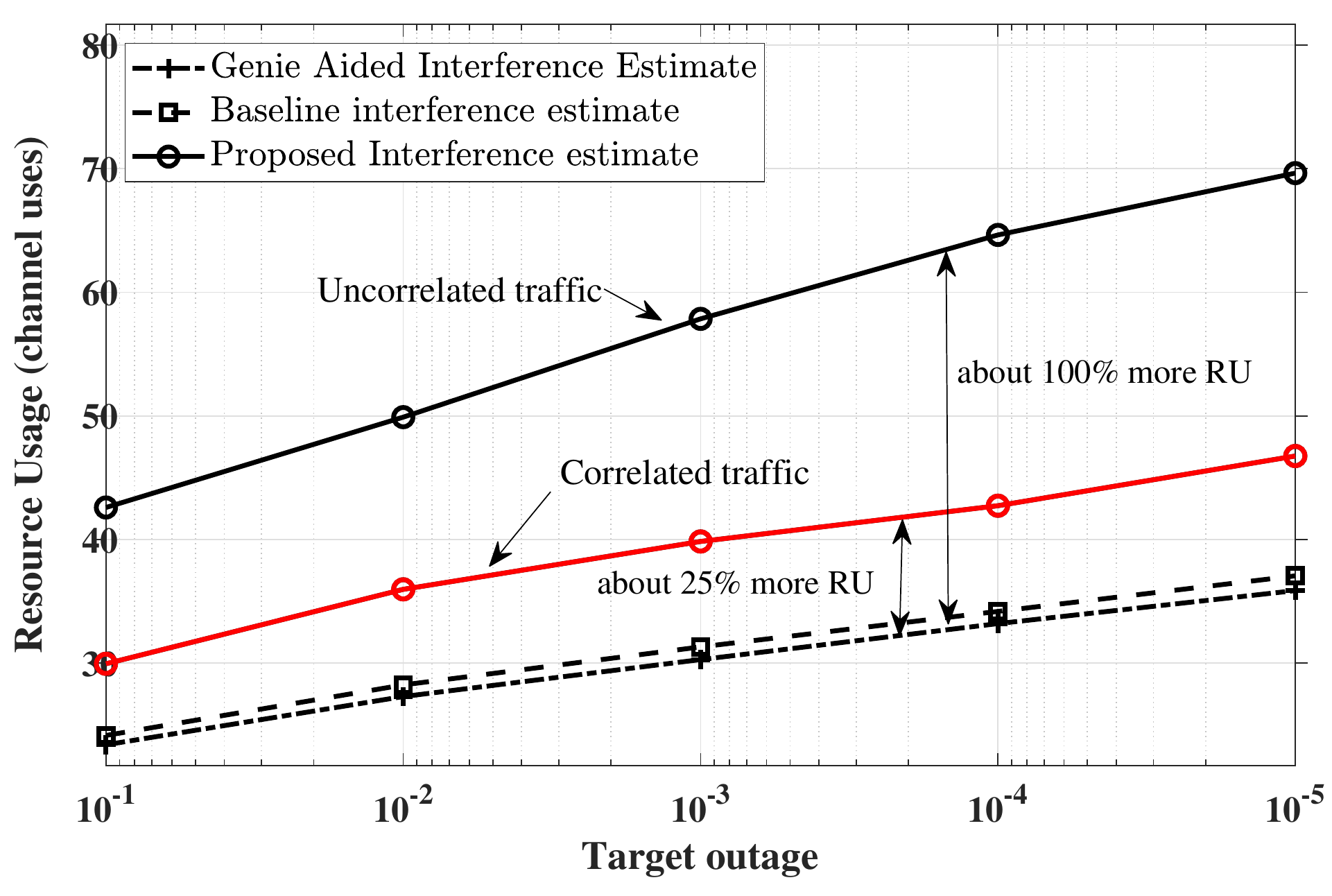}
    \caption{Resource usage in channel uses vs. target outage probability with correlated traffic, $\eta = 0.95$.}
    \label{fig:outage_vs_RU_corr}
\end{figure}

The outage probability as a function of the target outage considering (time) correlated and uncorrelated interference is presented in Fig.~\ref{fig:correlated}. Under these realistic assumptions, the proposed interference estimate based RRM scheme is found to always fulfill the target outage, whereas the baseline estimator performance is only slightly improved. Correlation induces greater dependence in the transition from one state to the next. Hence, estimating future interference values based on the state transition probability matrix results in greater accuracy compared to the uncorrelated traffic case. The corresponding RU, as shown in Fig.~\ref{fig:outage_vs_RU_corr}, is only about $25\%$ more than the optimum RU. This indicates that the proposed scheme is even more efficient when considering the realistic case of correlated traffic. The RU for correlated traffic is almost unchanged for both the baseline schemes, and hence not illustrated in the figure.

\section{Conclusions and Future Works}
\label{sec:conclusions}

URLLC mandates a departure from the conventional average-based resource management approach. Instead, intelligent and risk sensitive designs are needed to meet the stringent reliability and latency requirements while guaranteeing high resource efficiency. A novel interference prediction based RRM algorithm is proposed in this letter. The interference distribution is modeled as a discrete state space DTMC. Future interference states are predicted by utilizing the state transition probability matrix along with a risk sensitivity parameter. The reliability of the proposed scheme is found to be significantly better than that of the considered baseline RRM scheme. Alongside, the proposed scheme requires only slightly more resources than the optimum case. As next steps, we plan to investigate more resource efficient implementations of the proposed scheme. 


\begin{thebibliography}{10}
\providecommand{\url}[1]{#1}
\csname url@samestyle\endcsname
\providecommand{\newblock}{\relax}
\providecommand{\bibinfo}[2]{#2}
\providecommand{\BIBentrySTDinterwordspacing}{\spaceskip=0pt\relax}
\providecommand{\BIBentryALTinterwordstretchfactor}{4}
\providecommand{\BIBentryALTinterwordspacing}{\spaceskip=\fontdimen2\font plus
\BIBentryALTinterwordstretchfactor\fontdimen3\font minus
  \fontdimen4\font\relax}
\providecommand{\BIBforeignlanguage}[2]{{%
\expandafter\ifx\csname l@#1\endcsname\relax
\typeout{** WARNING: IEEEtran.bst: No hyphenation pattern has been}%
\typeout{** loaded for the language `#1'. Using the pattern for}%
\typeout{** the default language instead.}%
\else
\language=\csname l@#1\endcsname
\fi
#2}}
\providecommand{\BIBdecl}{\relax}
\BIBdecl

\bibitem{SWD+18}
J.~{Sachs} \emph{et~al.}, ``{5G} radio network design for ultra-reliable
  low-latency communication,'' \emph{IEEE Network}, vol.~32, no.~2, pp. 24--31,
  2018.

\bibitem{berardinelli2018_wirt}
G.~Berardinelli, N.~H. Mahmood, I.~Rodriguez, and P.~E. Mogensen, ``Beyond {5G}
  wireless {IRT} for {Industry} 4.0: Design principles and spectrum aspects,''
  in \emph{Proc. GC Workshops}, UAE, Dec. 2018.

\bibitem{Bennis.2018}
M.~{Bennis}, M.~{Debbah}, and H.~V. {Poor}, ``Ultrareliable and low-latency
  wireless communication: {Tail}, risk, and scale,'' \emph{Proceedings of the
  IEEE}, vol. 106, no.~10, pp. 1834--1853, Oct 2018.

\bibitem{MLP+19_MCA}
N.~H. Mahmood \emph{et~al.}, ``Multi-channel access solutions for {5G} new
  radio,'' in \emph{Proc. {IEEE} Wireless Communications and Networking
  Conference (WCNC) Workshops}, Marrakech, Morocco, Apr. 2019, pp. 1--6.

\bibitem{MTCwhitePaper2020}
------, \emph{White Paper on Critical and Massive Machine Type Communication
  towards {6G}}, ser. 6G Research Visions, nr. 11, N.~H. Mahmood \emph{et~al.},
  Eds.\hskip 1em plus 0.5em minus 0.4em\relax Oulu, Finland: University of
  Oulu, Jun. 2020.

\bibitem{Pocovi.2017}
G.~{Pocovi}, B.~{Soret}, K.~I. {Pedersen}, and P.~{Mogensen}, ``{MAC} layer
  enhancements for ultra-reliable low-latency communications in cellular
  networks,'' in \emph{2017 IEEE International Conference on Communications
  Workshops (ICC Workshops)}, May 2017, pp. 1005--1010.

\bibitem{LCH+20}
M.~{Li}, C.~{Chen}, C.~{Hua}, and X.~{Guan}, ``A learning-based pre-allocation
  scheme for low-latency access in industrial wireless networks,'' \emph{IEEE
  Transactions on Wireless Communications}, vol.~19, no.~1, pp. 650--664, Jan.
  2020.

\bibitem{MAB+19_grantFree}
N.~H. Mahmood \emph{et~al.}, ``Uplink grant-free access solutions for {URLLC}
  services in {5G} new radio,'' in \emph{Proc. 16th International Symposium on
  Wireless Communication Systems (ISWCS)}, Oulu, Finland, Aug. 2019, pp.
  607--612.

\bibitem{MPJ+16_oneStage}
N.~H. Mahmood, N.~Pratas, T.~H. Jacobsen, and P.~E. Mogensen, ``On the
  performance of one stage massive random access protocols in {5G} systems,''
  in \emph{Proc. 9th International Symposium on Turbo Codes and Iterative
  Information Processing, (ISTC)}, Brest, France, Sep. 2016, pp. 340--344.

\bibitem{Popovski.2018}
P.~{Popovski} \emph{et~al.}, ``Wireless access for ultra-reliable low-latency
  communication: {Principles} and building blocks,'' \emph{IEEE Network},
  vol.~32, no.~2, pp. 16--23, Mar. 2018.

\bibitem{polyanskiy_trIT2010}
Y.~Polyanskiy, H.~V. Poor, and S.~Verdu, ``Channel coding rate in the finite
  blocklength regime,'' \emph{IEEE Transactions on Information Theory},
  vol.~56, no.~5, pp. 2307--2359, May 2010.

\bibitem{AV_jsac2018}
A.~Anand and G.~de~Veciana, ``Resource allocation and {HARQ} optimization for
  {URLLC} traffic in {5G} wireless networks,'' \emph{IEEE Journal on Selected
  Areas in Communications}, vol.~36, no.~11, pp. 2411--2421, Nov. 2018.

\bibitem{LKD20_FBCsingle}
A.~{Lancho}, T.~{Koch}, and G.~{Durisi}, ``On single-antenna rayleigh
  block-fading channels at finite blocklength,'' \emph{IEEE Transactions on
  Information Theory}, vol.~66, no.~1, pp. 496--519, Jan. 2020.

\end{thebibliography}


\end{document}